\documentclass[final,3p,times,twocolumn]{elsarticle}
\usepackage{graphicx}
\usepackage{amssymb}
\usepackage{lineno}

\usepackage{hyperref}
\usepackage[frozencache]{minted}
\newcounter{bla}

\journal{Computer Physics Communications (ANL-HEP-148927)}
\date{December 4, 2018}

\begin{document}

\begin{frontmatter}



\title{ProIO: An Event-Based I/O Stream Format for Protobuf Messages}


\author[a,d]{D.~Blyth\corref{author}}
\author[c]{J.~Alcaraz}
\author[b]{S.~Binet}
\author[a]{S.V.~Chekanov}

\cortext[author] {Corresponding author.\\\textit{E-mail address:} dblyth@anl.gov}
\address[a]{HEP Division, Argonne National Laboratory, 9700 S. Cass Avenue, Argonne, IL 60439, USA}
\address[b]{Universit\'e Clermont Auvergne, CNRS/IN2P3, LPC, F-63000 Clermont-Ferrand, France}
\address[c]{Northern Illinois University, DeKalb, IL USA}
    \address[d]{Radiation Detection and Imaging (RDI) Technologies, LLC, 21215 N 36th Pl, Phoenix, AZ 85050, USA}

\begin{abstract}
ProIO is a new event-oriented streaming data format which utilizes Google's Protocol Buffers (protobuf) to be flexible and highly language-neutral.  The ProIO concept is described here along with its software implementations.  The performance of the ProIO concept for a dataset with Monte-Carlo event records used in high-energy physics was benchmarked and compared/contrasted with ROOT I/O.  Various combinations of general-purpose compression and variable-length integer encoding available in protobuf were used to investigate the relationship between I/O performance and size-on-disk in a few key scenarios.
\end{abstract}


\end{frontmatter}

%%\linenumbers

% Computer program descriptions should contain the following
% PROGRAM SUMMARY.

{\bf PROGRAM SUMMARY}
  %Delete as appropriate.

\begin{small}
\noindent
{\em Program Title: ProIO}                                          \\
{\em Licensing provisions: BSD 3-clause }                                   \\
{\em Programming language: Python, Go, C++, Java}                                   \\
%{\em Supplementary material:}                                 \\
  % Fill in if necessary, otherwise leave out.
{\em Nature of problem:}\\
In high-energy and nuclear physics (HEP and NP), Google's Protocol Buffers
    (protobufs) can be a useful tool for the persistence of data.  However,
    protobufs are not well-suited for describing large, rich datasets.
    Additionally, features such as direct event access, lazy event decoding,
    general-purpose compression, and self-description are features that are
    important to HEP and NP, but that are missing from protobuf. \\
{\em Solution method:}\\
The solution adopted here is to describe and implement a streaming format for
    wrapping protobufs in an event structure.  This solution requires small
    (typically less than 1000 lines of code) implementations of the format in
    the desired programming languages.  With this approach, most of the I/O
    heavy lifting is done by the protobufs, and ProIO adds the encessary
    physics-oriented features. \\

\end{small}

\section{Introduction}
\label{SecIntroduction}
Google's Protocol Buffers (protobufs) \cite{pb} are an attractive tool for the persistence of data due to their ability to provide efficient serial encoding of user-defined data structures.  Variable-length encoding for integers (varints) is built into protobufs, and can be used as a physics-motivated compression technique that in some scenarios can lead to significant improvements in comparison to the application of general-purpose compression alone \cite{chekanov2014promc}.  The use of field identifiers in protobufs also leads to compression of sparse data and a well-defined mechanism for ensuring forwards and backwards compatibility (i.e. fields can be added or removed as a data model evolves without breaking compatibility with older versions of the data model).

\begin{figure*}
\centering
\includegraphics[width=4.5in]{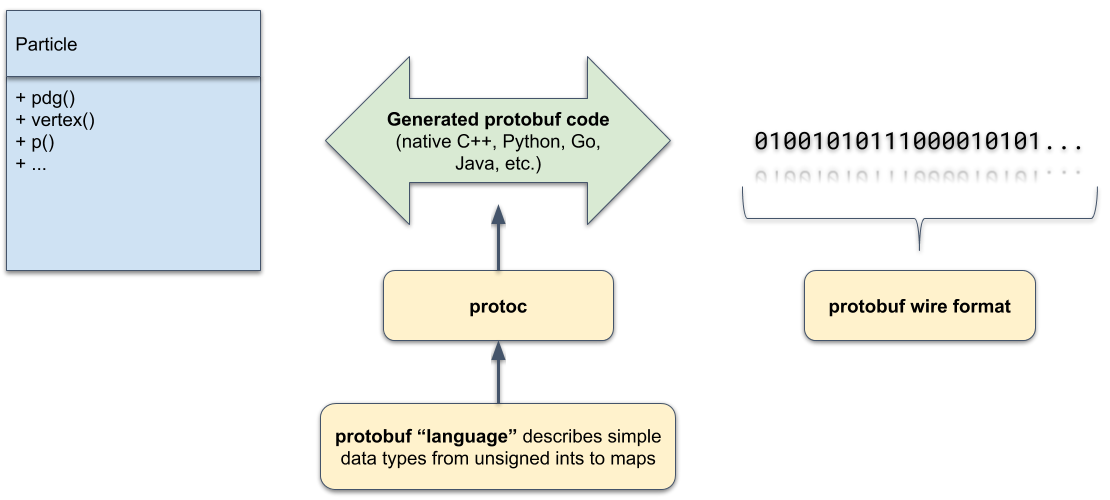}
\caption[Protocol Buffers]{\textbf{Protocol Buffers} - Google's Protocol Buffers (protobufs) provide a performant and language-neutral method of serializing data structures.  Data structures are described in a programming-language neutral protobuf "language", and language-specific code is generated automatically.}
\label{figProtobuf}
\end{figure*}

The way protobufs are created is also unique and powerful.  A protobuf language specification is used to describe data structures in a way that is programming-language neutral, and a protobuf ``compiler'' (protoc) is used to generate code in a desired programming language (see Figure \ref{figProtobuf}).  This generated code creates representations of the data structures in the desired programming language, and these representations are able to self-serialize and deserialize.  Not only does this process produce a performant method of data serialization that can be optimized at compile time, but it also allows the serialization to be implemented in any language from a single source.  In fact, industry-supported languages are numerous, and include most commonly-used modern languages.

On the other hand, protobufs suffer for being low-level and lacking in features that are important in high-energy and nuclear physics (HEP and NP) \cite{blomer}.  A single protobuf message can in principle define a file format itself, but this file would then have to be serialized or deserialized all at once, typically on a single thread.  For protobufs to be used to describe large numbers of rich events, structure must be added on top of protobufs that implements features like direct event access, lazy event decoding, additional general-purpose compression, and data self-description.

Previous work on utilizing protobufs for persistence of high-energy physics data gave rise to the ProMC file format \cite{chekanov2014promc}, which was designed for archiving event records from Monte-Carlo event generators \cite{Chekanov:2014fga}.  The main motivation for the ProMC format was in reduction of file sizes by taking advantage of variable-length encoding.  The format proposed in this paper also inherently takes advantage of this protobuf feature.  In contrast to ProMC, it is designed from the ground up to be more general in application and to allow greater flexibility in the use of compression algorithms.  ProIO is intended to be used with arbitrary event structure, such as Monte-Carlo events after detector simulation and event reconstruction, as well as experimental data.

ProIO is defined as an event-based stream format for protobuf messages.  For one, this means that it is focused on being stream compatible, and is not exclusively an archival file format.  For example, ProIO can be used as a serialization layer between processes without writing to disk.  Secondly, this definition indicates that ProIO is not inherently columnar, but instead event- or row-oriented.  Data are written and read as events that are effectively containers for arbitrary protobuf message entries, where the message entries represent parts of an event, e.g. detector hits.  However, this is not to say that entire events must be deserialized at once, and we will come back to this in discussion about lazy decoding.  Finally, note that ProIO itself is a format, and not an implementation of the format.  Libraries for reading and writing the ProIO format are already written in several programming languages, and command-line tools exist for inspecting and manipulating streams.  This collection of libraries and tools will also be outlined in this document.

\section{The Format}
\begin{figure}
\centering
\includegraphics[width=3in]{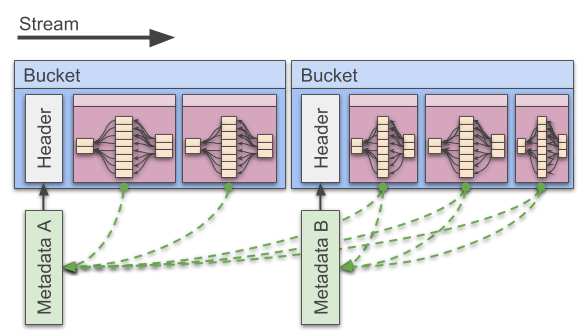}
\caption[ProIO Stream]{
\textbf{ProIO Stream} - A ProIO stream is simply a sequence of events described in protobuf format that are grouped into what are called buckets.  Bucket headers describe what general-purpose compression (such as GZIP or LZ4) is applied to the bucket, as well as user-defined metadata (or slow data) that are synchronized with the event stream.  Additionally, bucket headers provide information necessary to decode the data stream if the data model is unknown at compile time.
}
\label{figProIOStream}
\end{figure}

\subsection{Buckets}
ProIO streams are segmented into what are called buckets (see Figure \ref{figProIOStream}).  A bucket is a collection of consecutive events that are optionally compressed together, and each bucket has a header.  Buckets are intended to be O(MB)-sized, i.e. large enough for efficient compression and also much larger than the header data.  On disk, this also translates to bucket headers occupying a very small fraction of the total number of disk sectors used by the file.  This is important for reasonably fast direct access of events with a cold disk cache, since ProIO streams do not contain global locations of events.

\begin{figure}
\centering
\includegraphics[width=3in]{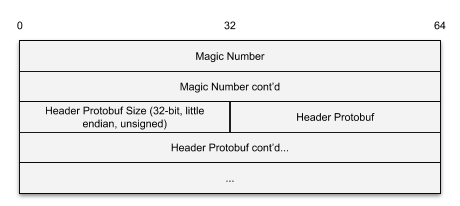}
\caption[Bucket Header]{\textbf{Bucket Header} - Bucket headers primarily contain magic numbers for synchronization and protobuf messages that contain details about the buckets.}
\label{figBucketHeader}
\end{figure}

\paragraph{Header}
Each bucket has a header that describes the bucket (see Figure \ref{figBucketHeader}). This header is also an opportunity to resynchronize/recover the stream so that in principle corruption within a bucket is isolated. This synchronization is achieved with a magic number. This is a special sequence of 128 bits that identifies the start of the bucket header, and also serves to identify ProIO stream compatibility. The magic number is \texttt{0xe1}, \texttt{0xc1}, followed by 14 bytes of zeros. If ProIO stream compatibility is ever broken, this magic number will be changed. Following the magic number is an unsigned 32-bit little-endian integer that states the size of the remaining header in bytes.  The remaining header is in protobuf wire format and described in protobuf language.  The exact protobuf structure for the header is described in \cite{proioRepo}.

\paragraph{Metadata}
The bucket header protobuf structure can contain metadata entries which are key-value pairs with strings as keys and byte arrays as values. These metadata are inserted into the stream at a certain point via a bucket header, and are to be associated with all events that follow until new metadata entries with the same keys as older entries override the older entries, or until the stream ends (see Figure \ref{figProIOStream}).

\paragraph{``FileDescriptorProto''s}
Bucket headers can store serialized, uncompressed ``FileDescriptorProto''s for describing protobuf messages.  A ``FileDescriptorProto'' is a protobuf message included in the protobuf libraries that is used to describe the language-neutral protobuf structure.  This is a critical ingredient of self description for ProIO. ProIO streams are required to contain bucket headers with serialized ``FileDescriptorProto''s before the corresponding message types appear in the stream.

\paragraph{Bucket Contents}
The contents of a bucket immediately follow the bucket header, and consist of an optionally-compressed set of consecutive event protobuf messages (described below and in protobuf language in \cite{proioRepo}), each preceded by an unsigned, 32-bit value stating the number of bytes to grab for that event message.

\paragraph{Compression}
If the bucket is compressed as indicated by the protobuf header data, the entire bucket is compressed together, and must be uncompressed prior to deserializing events.  Note, however, that the bucket does not necessarily need to be uncompressed all at once, and events may be accessed directly by uncompressing the bucket from the beginning of the bucket to the end of the event.  For LZ4 compression, the format is that of an LZ4 ``frame'', and in general each compression algorithm is written in its corresponding high-level format.  ProIO implementations are not required to support all compression types, but should handle unknown compression types gracefully.

\subsection{Events}
\begin{figure}
\centering
\includegraphics[width=3in]{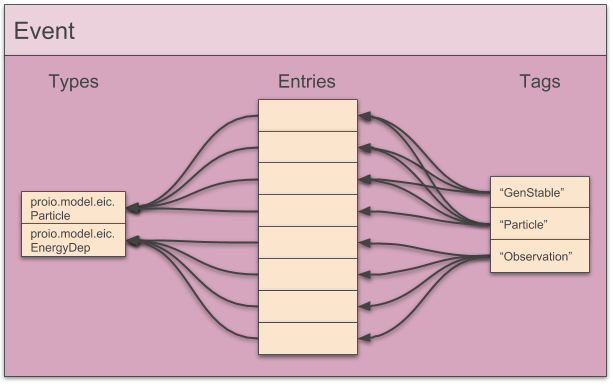}
\caption[ProIO Event]{\textbf{ProIO Event} - Each event carries a set of entries, types, and tags.  The actual data of the stream are stored as event entries and organized by tags.  Each event carries information about the entry types, i.e. what protobuf messages were used to write the entry data.}
\label{figProIOEvent}
\end{figure}

ProIO events are self-serializing containers for arbitrary protobuf messages. In a ProIO stream, events themselves are represented by protobuf messages, and the protobuf structure is described in \cite{proioRepo}.  As shown in Figure \ref{figProIOEvent}, an event has three main attributes: entries, types, and tags.

\paragraph{Entries and types}
The main purpose of a ProIO event is to store entries, which are arbitrary protobuf messages.  When an entry is added to an event, a string representing the entry message type is also added to the event's list of types if it is not already there.  The entry is stored in its serialized form as an array of bytes, which are associated with a pointer to its type string.  The protobuf description of the event does not know the structure of the entry, only that it is an array of bytes.  When an event is read, the code responsible for reading (a library implementation) looks up the entry's associated type and uses the appropriate precompiled protobuf code to deserialize the entry, or a dynamic protobuf message representation is used which takes information from the stream's accumulated ``FileDescriptorProto''s.  This type specification is the other critical ingredient of ProIO self-description.

\paragraph{Tags}
Each event also stores a list of tags for organizing the entries.  A tag is a mapping from a human-readable string to a list of entries.  Any number of tags may point to a particular entry.

\section{Implementations}
There are currently native ProIO library implementations in C++, Python, Go, and Java (though Java is read-only at the time of this writing).  The home of these projects is the ``proio-org'' GitHub organization which serves to organize the four code bases along with the language-independent files that describe the header and event structure (\cite{proioRepo}).

\subsection{Library Usage}
In each implementation, there is a stream reader type, a stream writer type, and an event type.  The reader acts as an iterator over events in a stream, handles general-purpose decompression of bucket contents, and manages the association of events with metadata as well as the available entry type descriptors (from the bucket headers' ``FileDescriptorProtos'') collected from the stream.  The writer facilitates pushing events into a stream and injecting metadata and entry type descriptors for stream self description, and handles general-purpose compression of bucket contents.  Finally, the event type is a language-specific handle for manipulating the event data, e.g. adding and retrieving entries, assigning tags, etc.

Adding an entry to an event typically involves
\begin{enumerate}
    \item creating an entry protobuf message and assigning its data members,
    \item calling an add-entry method on the event handle and assigning at least one human-readable tag to the entry,
    \item and holding onto the numeric identifier returned by the add-entry method for further persistent referencing of the entry (such as by another entry).
\end{enumerate}
An example of writing an event in Python (py-proio \cite{py-proioRepo}) is given below.
\begin{minted}[linenos]{python}
import proio
import proio.model.example as model

with proio.Writer('file.proio') as w:
  ev = proio.Event()
  part = model.Particle()
  part.pdg = 11
  id = ev.add_entry('Stable', part)
  w.push(ev)
\end{minted}

Retrieving entries from an event typically involves
\begin{enumerate}
    \item getting a list of numeric entry identifiers for a particular tag,
    \item calling a get-entry method for each identifier to obtain the deserialized entry,
    \item and casting the returned entry into a known type in statically-typed languages.
\end{enumerate}
An example of reading events in Python is given below.
\begin{minted}[linenos]{python}
import proio

with proio.Reader('file.proio') as r:
  for ev in r:
    for id in ev.tagged_entries('Stable'):
      print(ev.get_entry(id))
\end{minted}

\paragraph{In ROOT}
In the C++ implementation (cpp-proio \cite{cpp-proioRepo}) installation, ROOT ``rootmap'' files are included.  Because of this, simply using ``\#include'' statements on cpp-proio headers will automatically link the installed libraries.  For example, the following code can be executed in ROOT's Cling environment following proper installation of cpp-proio:
\begin{minted}[linenos]{cpp}
#include "proio/reader.h"
#include "proio/model/example/example.pb.h"
namespace model = proio::model::example;

proio::Reader r("file.proio");
proio::Event ev;
while (r.Next(&ev)) {
  cout << ev.String() << endl;
}
\end{minted}

For more detailed examples in Python, Go, C++, and Java, see \ref{appPythonExample}, \ref{appGoExample}, \ref{appCPPExample}, and \ref{appJavaExample}, respectively.

\subsection{Tools}
A set of command-line tools are available for inspecting and manipulating ProIO streams.  These tools are written in Go for a powerful combination of portability and speed, and can be found in \cite{go-proioRepo}.  Pre-compiled binary versions of the tools for Linux, Mac OS, and Windows can be found under the releases tab at the cited URL.
\begin{itemize}
    \item \emph{proio-summary}: scans a stream until the end and gives a brief summary of the contents
    \item \emph{proio-ls}: dumps string representations of entry data that are tagged with specific tags and/or in a specific event
    \item \emph{proio-strip}: removes entry data from the stream matching (or not matching) specific combinations of tags - An example use case is to separate a stream of data into two parts for separate processing.
    \item \emph{lcio2proio}: converts data from the LCIO format \cite{gaede2003lcio} to ProIO using Go-HEP \cite{binet2018go} for LCIO input
\end{itemize}

A GUI browser implemented in Java has also been developed for easier browsing of data, and is contained in \cite{java-proioRepo}.

\section{Benchmarks}
In order to assess the performance of ProIO streams for use in HEP and NP, we use ROOT as a standard for comparison.  We focus on a particular use case that is a reasonable application of both ProIO and ROOT \cite{root}.  This is the case of Monte-Carlo generator output with particles described with each attribute as an array where the index of the array refers to a particular particle.  We refer to this format as a packed particle format.

We used Pythia8 \cite{Sjostrand:2007gs} to generate proton-proton collision events at 13 TeV.  Each generated event contains a dynamic number of particles.  In ROOT, each particle attribute was stored as a separate branch, with one branch dedicated to specifying the number of particles for each event.  The attribute branches each have a single C-style array leaf.  In ProIO, each event stores a single protobuf message representing a list of particles (this is the `PackedParticles` message type of the `example` data model in \cite{proioRepo}).  This message has a repeated, packed field for each particle attribute.  For the ProIO implementation, we used cpp-proio \cite{cpp-proioRepo}.

16000 events were generated, each with an average of $\sim1500$ particles.  This dataset was reencoded with each available compression algorithm applied at a high compression level.  For ROOT, the encodings are uncompressed, LZ4-9, ZLIB-7, and LZMA-9, where the number indicates the compression level.  For cpp-proio, the configurations are uncompressed, LZ4-9, and GZIP-7.  For each configuration, we have generated lossless, floating-point precision representations, as well as lossy, fixed-point precision representations using integers where applicable.  For the fixed-point precision representations, the least-significant digit was in units of 10~keV for energy, and 1~micron for spacetime coordinates.  This is to examine the efficiency of varint compression in protobufs compared to the use of fixed-length integer encoding in ROOT with general-purpose compression applied on top.  The idea here is not so much to compare the most compact possible representations, but to examine I/O overhead as a function of size-on-disk.

These benchmarks were performed using Linux 4.18 on an Intel Core i5 6400 CPU with an NVMe solid-state drive.  The benchmarks were containerized using Docker, and all code including Dockerfile source can be found in \cite{bench-repo}, enabling precise replication of results shown here.

\subsection{Reencoding}
\begin{figure*}
\centering
\includegraphics[width=3in]{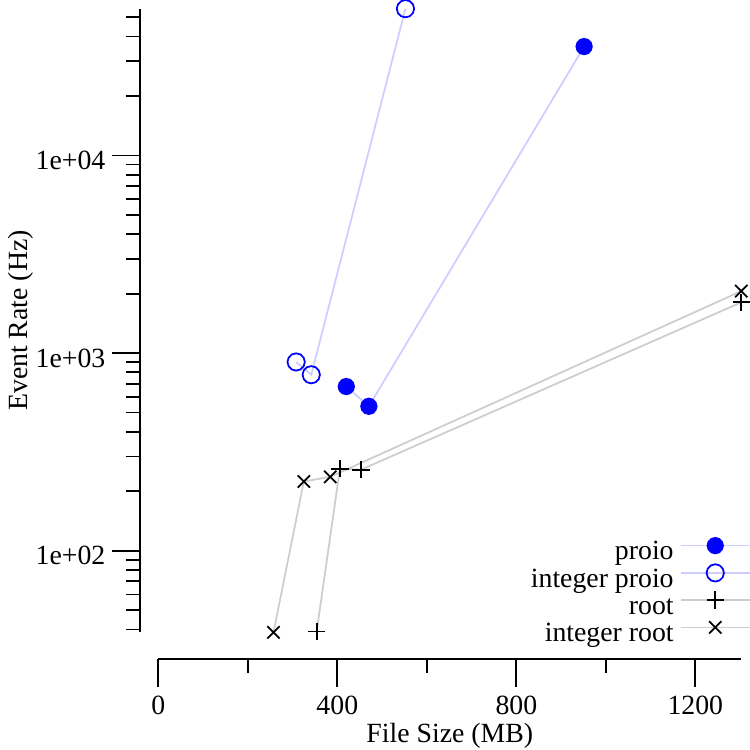}
\includegraphics[width=3in]{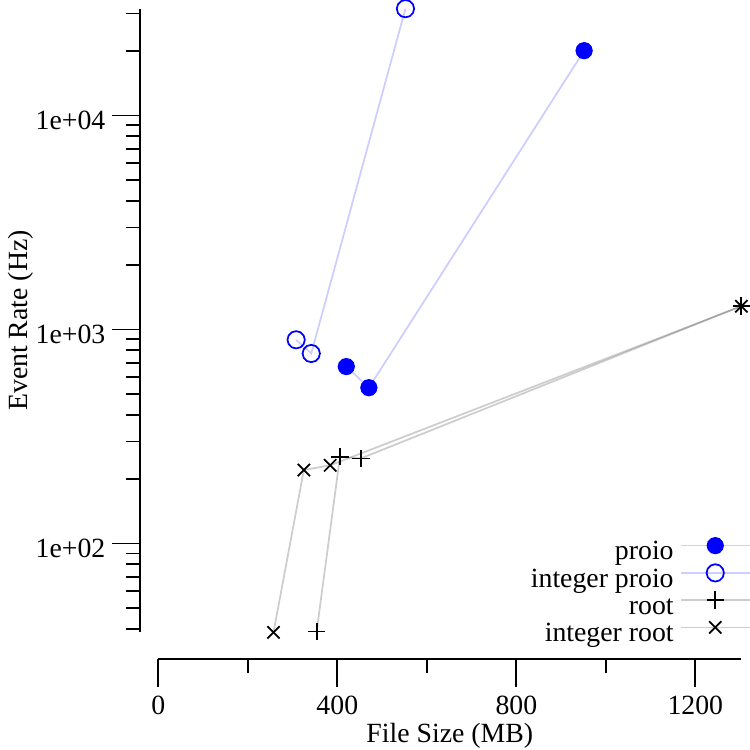}
\caption[Reencode Event Rate]{
\textbf{Reencode Event Rate} - The event rates achieved in the reencoding process as a function of the resulting file sizes are shown.  In all cases, the source data were stored in memory via tmpfs.  On the \textbf{LEFT}, the event rates are measured using the total userspace CPU time used.  On the \textbf{RIGHT}, the rates are measured using the total wall time.
}
\label{figReencodeRate}
\end{figure*}

In order to produce the various compression configurations, the ``uncompressed'' data were read and rewritten (reencoded) with the appropriate compression settings.  It is useful to compare the performance of this process between ProIO and ROOT in order to highlight some fundamental differences.  In Figure \ref{figReencodeRate}, reencoding speed in terms of event rate is shown as a function of the resulting file size.  Note that each point represents one of the compression configurations described above, but only the resulting size-on-disk is indicated, rather than the algorithm.  ProIO and ROOT files are distinguished from one another, and we additionally distinguish between the lossless, floating-point precision data and the lossy, fixed-point precision data.  Note also that even for the non-integer, ``uncompressed'' cases, the file sizes are different between ROOT and ProIO because of ProIO varint compression of the parts of the data that are already naturally integers.

For these measurements, the source data were stored in memory via a tmpfs filesystem in order to remove storage read overhead, and written to an NVMe SDD.  Therefore, the measured rates primarily indicate the speed of the decoding and subsequent encoding process.  Rates measured using both userspace CPU time and wall time are given.  These rates are very similar because of write caching.  Write caching was left enabled for these measurements since disabling it may introduce a bias that is not meaningful in a realistic scenario.

As seen in Figure \ref{figReencodeRate}, the rates for ProIO reencoding are much higher than for ROOT.  A significant reason for this is the ProIO lazy decoding mechanism.  Because of lazy decoding, the actual physics data are not decoded and reencoded in the process of reencoding the file.  The data are read and rewritten largely untouched.  From profiling data, we know that for the case of the integer ProIO reencoding process with LZ4 compression applied, the code spends approximately 98\% of its time applying the `LZ4HC` compression algorithm.  Interestingly, this number only drops to about 88\% for the integer ROOT encoding, but about 3x the time is spent applying the algorithm.  This extra time spent in the `LZ4HC` algorithm with the ROOT data makes sense given the much smaller intrinsic size of the varint-encoded ProIO data.

\subsection{Sequential Access}
\begin{figure*}
\centering
\includegraphics[width=3in]{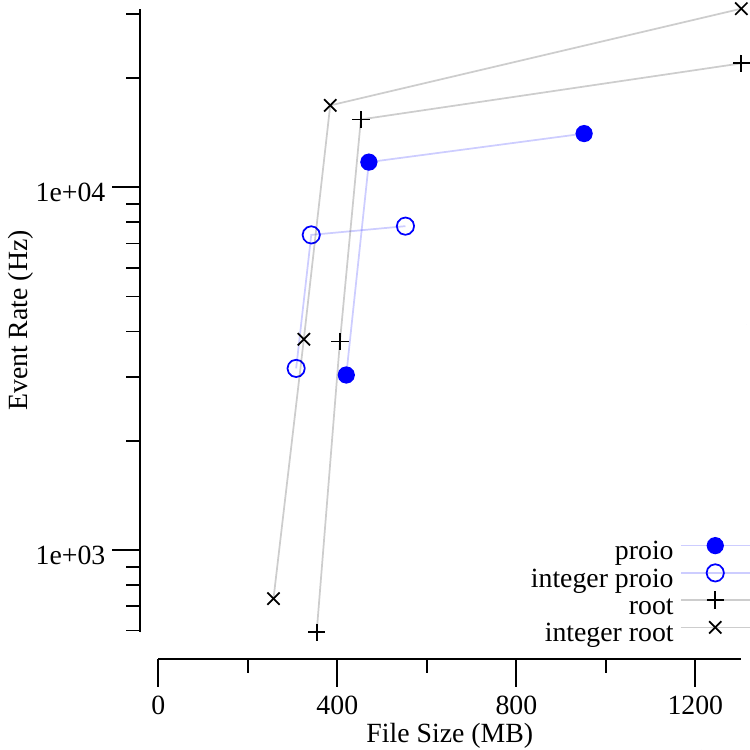}
\includegraphics[width=3in]{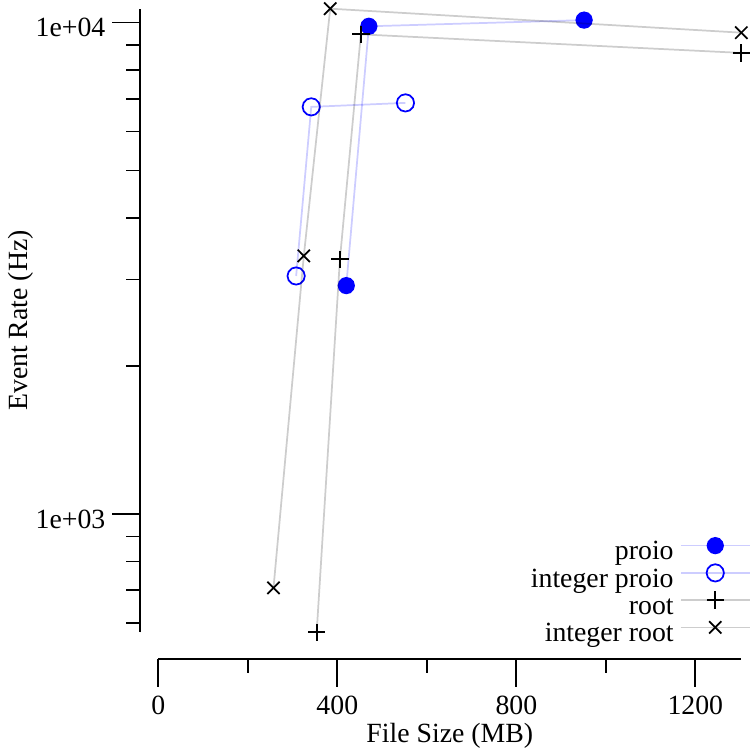}
\caption[Sequential Decode Event Rate]{
\textbf{Sequential Decode Event Rate} - The event rates achieved in the sequential decoding process as a function of the source file sizes are shown.  On the \textbf{LEFT}, the event rates are measured using the total userspace CPU time used.  On the \textbf{RIGHT}, the rates are measured using the total wall time.
}
\label{figSequentialDecodeRate}
\end{figure*}

One of the most important measurements to make is simple sequential read rates of the entire event data.  For these measurements, the files are stored on an NVMe SSD and read with a cold filesystem cache.  Rates measured using both userspace CPU time and wall time are given in Figure \ref{figSequentialDecodeRate}.

Though we would not expect major differences between ProIO and ROOT in this benchmark, it is important for informing the choice of compression as well as to assess whether or not there is improvement in CPU-time efficiency from varint encoding.  As seen on the left of Figure \ref{figSequentialDecodeRate}, there is no CPU-time advantage to using varint encoding in sequential reads.  In fact, for this dataset, there appears to be no practical performance difference between the two formats.  In all ProIO data points, there is some degree of varint compression overhead.  The varint compression appears to reduce file sizes at a cost that is comparable to general-purpose compression.

\paragraph{Columnar Access}
Of course, a common use case in ROOT is to access a particular column of the dataset.  This is not directly benchmarked here because the comparison is highly dependent on the particular scenario.  It suffices to say that ProIO tends to suffer in this type of access in cases where storage read speeds are a limiting factor.  In ProIO, all event entries are read from storage together.  However, entries are only decoded upon access (i.e., the the event is lazily decoded).  This means that data that are not part of a ``column'' do not contribute to the decoding (including varint decoding) overhead.  Additionally, ProIO metadata can be another means of mapping data into columns.  By filtering events by metadata in order to map into columns, ProIO files are expected to have similar performance to ROOT in columnar access.

\paragraph{Dynamic Decoding}
\begin{figure}
\centering
\includegraphics[width=3in]{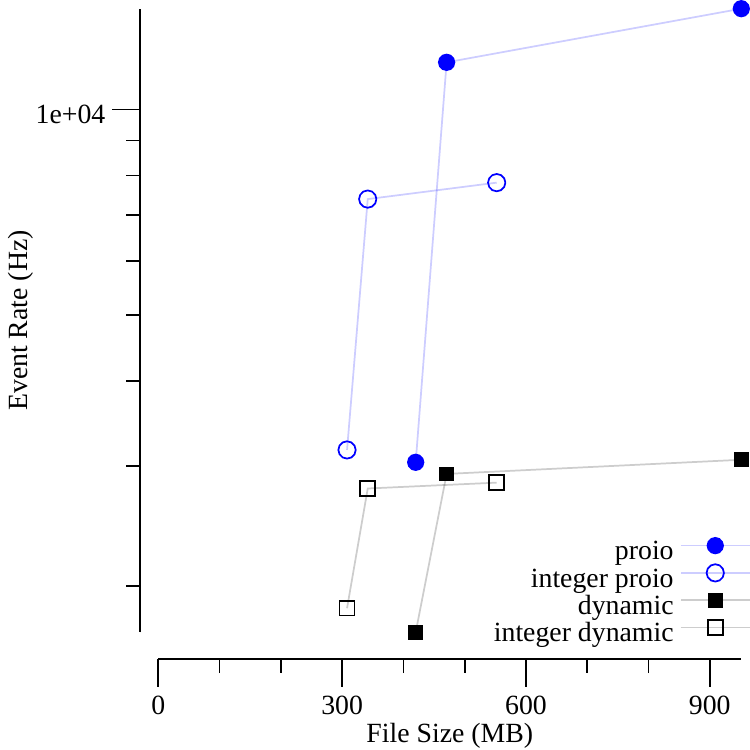}
\caption[Dynamic Decode Event Rate]{\textbf{
Dynamic Decode Event Rate} - Event rates achieved in the sequential decoding process as a function of the source file sizes are shown.  The rates are measured using the total userspace CPU time used.  The same rates for ProIO data from Figure \ref{figSequentialDecodeRate} are shown along with rates for dynamic decoding of the data structures that are only known at run time.
}
\label{figDynamicDecodeRate}
\end{figure}

An important feature of ProIO is the self-descriptive nature of streams.  The preferred method of reading streams is to have precompiled and optimized code for a particular data model.  However, ProIO claims to be self-descriptive because of the fact that data can also be decoded without knowledge of the data model at compile time.  The cpp-proio library takes advantage of the `DynamicMessage` feature of the C++ protobuf library and the ``FileDescriptorProto''s provided by the ProIO stream in order to dynamically decode the data.  The performance of this mode of operation is shown in Figure \ref{figDynamicDecodeRate}.

As seen in Figure \ref{figDynamicDecodeRate}, the performance penalty from dynamically decoding the data is significant.  It is worth noting, however, that for future work there is a reasonably straightforward path to introducing just-in-time (JIT) compilation of code for decoding unknown data.  The same ``FileDescriptorProto''s that are currently used with the `DynamicMessage` protobuf feature can be used to generate code on the fly to be JIT compiled.  This code generation process can tap in directly to the existing process of generating protobuf code.

\subsection{Random Access}
\begin{figure}
\centering
\includegraphics[width=3in]{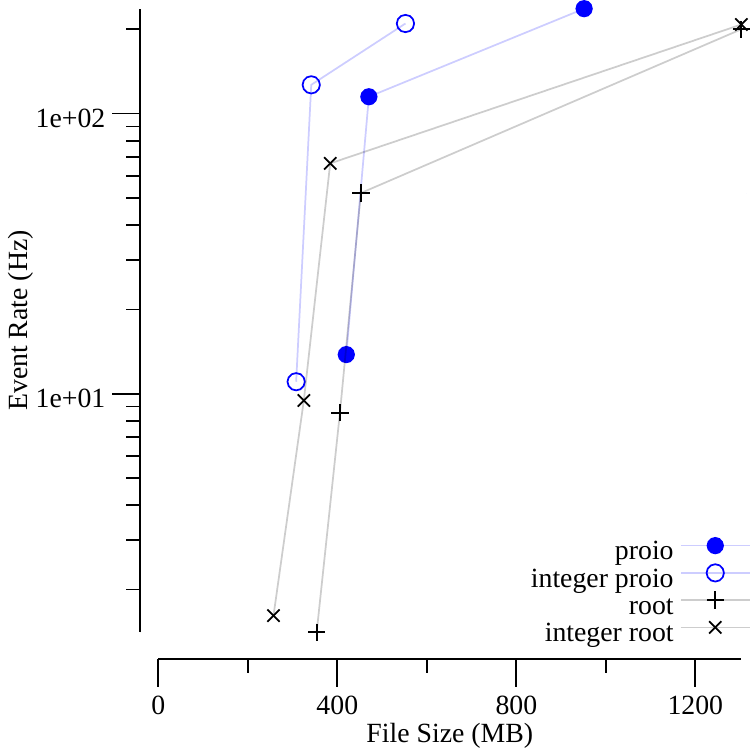}
\caption[Random Access Event Rate]{
\textbf{Random Access Event Rate} - The event rates achieved in random event access as a function of the source file sizes are shown.  The rates are measured using the total wall time in a scenario that simulates fully cached files.  It is important to note that the scales of these values are highly dependent on bucket/buffer sizes, which can be tuned.
}
\label{figRandomAccessRate}
\end{figure}

The performance of general-purpose compression has been found above to be on par with varint compression for sequential access of whole events.  However, the fact that varint compression is a kind of local or intrinsic compression to the encoding can provide a performance benefit in other scenarios.  As touched on above, one of these scenarios is the partial decoding of an event.  Another scenario is in the direct, or random access of an event (Figure \ref{figRandomAccessRate}).

In direct access of an event, any general purpose compression that is applied to a bucket/buffer introduces extra overhead from the decompression of surrounding data.  Varint compression does not have this drawback.  To examine the effect that this has, Figure \ref{figRandomAccessRate} gives rates for decoding entire events chosen at random throughout the entire dataset.  Since caching can influence this heavily, the measurements were made from files stored in memory via tmpfs.  The results show that the protobuf varint compression used in the ProIO files in fact does improve the rate of random access.  The sizes of the buckets/buffers can scale these data points to a large extent, and this is a tunable parameter.  However, the defaults for both cpp-proio and ROOT happen to provide similar performance here.

\section{Summary and Discussion}
In this document, the ProIO format and software written for implementing this format have been described.  This serves to help define the format as well as to help the reader become familiar with how to use ProIO in a project.  Since ProIO is not inherently a columnar data format, there is a degree of orthogonality with respect to ROOT I/O.  Additionally, ProIO was developed with stream compatibility in mind, which further separates it.

The performance of the ProIO concept has been benchmarked in comparison with ROOT I/O.  We have shown that ProIO performs well when applied to a benchmark dataset based on Monte Carlo event records.  The benchmarks were designed to be as fair as possible in the comparison between ProIO and ROOT I/O. Additionally, it should be pointed out that the comparison made in this document is not to suggest ProIO as a replacement for ROOT I/O.  Instead, it aims to provide some contrast to what is a highly ubiquitous method of storing data in HEP and NP.

Ultimately, the main motivator of and benefit to the development and use of ProIO is language neutrality.  It is becoming more and more clear that there is real benefit in bringing HEP and NP analysis to new programming languages that support highly portable and transparent code with fast development cycles.  While some efforts to bring native ROOT I/O capability to languages other than C++ (in Go \cite{binet2018go} and Python \cite{uproot}, e.g.) are quite successful, there is an inherent barrier to language neutrality introduced by defining a complex file format in a particular programming language.  Leveraging Protocol Buffers, where the structure of the data is described in a way that is purposefully language-neutral, has the potential to remove much of this barrier.  Additionally, Protocol Buffers are widely used and supported by industry in many languages already, making their use in HEP and NP a relatively small task.

Another benefit of ProIO is due to its focus on streaming data.  In the Electron-Ion Collider (EIC) Software Consortium (ESC), it has been found that collaboration between institutions on common software projects may be enhanced by the development of a common data model at various points in the process of simulation, reconstruction, and analysis of EIC physics events \cite{diefenthaler-talk}.  These common data model points encourage interoperability between software, making the replacement of part of the software chain easier.  However, without a streaming data format this kind of practice could quickly make I/O a bottleneck.  Using ProIO streams to mediate the flow of data in the simulation, reconstruction, and analysis software is currently in practice at Argonne National Laboratory.

\section*{Acknowledgments}
The submitted manuscript has been created by UChicago Argonne, LLC, Operator of Argonne National Laboratory (“Argonne”). Argonne, a U.S. Department of Energy Office of Science laboratory, is operated under Contract No. DE-AC02-06CH11357. The U.S. Government retains for itself, and others acting on its behalf, a paid-up nonexclusive, irrevocable worldwide license in said article to reproduce, prepare derivative works, distribute copies to the public, and perform publicly and display publicly, by or on behalf of the Government.  The Department of Energy will provide public access to these results of federally sponsored research in accordance with the DOE Public Access Plan. \url{http://energy.gov/downloads/doe-public-access-plan}. Part of Argonne National Laboratory’s work was funded by the U.S. Department of Energy, Office of High Energy Physics under contract DE-AC02-06CH11357.  The work of D. Blyth was supported by an ANL LDRD project 2017-058.  The work of J. Alcaraz was supported by the EIC RD14 project.

D.~Blyth and S.V.~Chekanov would also like to thank Alexander Kiselev for his constructive feedback on the conceptual development of ProIO, and Jakob Blomer for his help with ROOT I/O comparison.

%% References


\bibliographystyle{cpc}
\bibliography{proio}






% \end{thebibliography}

\appendix
\onecolumn

\section{Python Example}
\label{appPythonExample}
The following example uses py-proio \cite{py-proioRepo} to write and read in the ProIO format.
\paragraph{Write Particles to File or Stream}
\begin{minted}[linenos]{python}
import proio
# import data model protobufs (this example model happens to be included in the
# proio package, as are a few others)
import proio.model.example as model

# open a stream writer scope that ouputs to 'file.proio'
with proio.Writer('file.proio') as writer:
    # label all future events as 'run' = 'jpsi0', until this metadata key is
    # overwritten
    writer.push_metadata('run', b'jpsi0')

    event = proio.Event()
    
    # add particles to the empty event and hold onto their assigned identifiers
    parent = model.Particle()
    parent.pdg = 443
    parent.p.x = 1
    parent.mass = 3.097
    parent_id = event.add_entry('Particle', parent)

    child1 = model.Particle()
    child1.pdg = 11
    child1.vertex.x = 0.5
    child1.mass = 0.000511
    child1.charge = -3

    child2 = model.Particle()
    child2.pdg = -11
    child2.vertex.x = 0.5
    child2.mass = 0.000511
    child2.charge = 3

    child_ids = event.add_entries('Particle', child1, child2)
    # tag the stable particles with a special tag
    for ID in child_ids:
        event.tag_entry(ID, 'GenStable')

    # use assigned identifiers to reference particle lineage
    parent.child.extend(child_ids)
    child1.parent.append(parent_id)
    child2.parent.append(parent_id)

    print(event)
    # push the event into the stream (save to file)
    writer.push(event)
\end{minted}

\paragraph{Read and Iterate Particles}
\begin{minted}[linenos]{python}
import proio
# no need to import data model when reading since streams are self-descriptive
# and dynamic deserialization is done seamlessly in python

# open a stream reader scope that inputs from 'file.proio'
with proio.Reader('file.proio') as reader:
    # loop over, and enumerate events in stream (in this case a file)
    for evnum, event in enumerate(reader):
        print('EVENT: %d, RUN: %s' % (evnum, event.metadata['run'].decode("utf-8")))
        # get entry identifiers tagged as 'Particle'
        parts = event.tagged_entries('Particle')
        print('%i particle(s)...' % len(parts))
        # loop over particle identifiers
        for i, id in enumerate(parts):
            # deserialize particle entry
            part = event.get_entry(id)
            print('%i. PDG Code: %i' % (i, part.pdg))

            print('  %i children...' % len(part.child))
            for j, id in enumerate(part.child):
                print('  %i. PDG Code: %i' % (j, event.get_entry(part.child[j]).pdg))
\end{minted}

\section{Go Example}
\label{appGoExample}
The following example uses go-proio \cite{go-proioRepo} to write and read in the ProIO format.
\paragraph{Write Particles to File or Stream}
\begin{minted}[linenos]{go}
package main

import (
        "fmt"
        "log"

        "github.com/proio-org/go-proio"
        // import data model protobufs (this example model happens to be included
        // in the generated go-proio-pb package, as are a few others)
        model "github.com/proio-org/go-proio-pb/model/example"
)

func main() {
        // create a writer for outputting to "file.proio"
        writer, err := proio.Create("file.proio")
        if err != nil {
                log.Fatal(err)
        }
        defer writer.Close()

        // label all future events as "run" = "jpsi0", until this metadata key is
        // overwritten
        writer.PushMetadata("run", []byte("jpsi0"))

        event := proio.NewEvent()

        // add particles to the empty event and hold onto their assigned
        // identifiers
        parent := &model.Particle{
                Pdg:    443,
                Vertex: &model.XYZTF{X: 1},
                Mass:   3.097,
        }
        pid := event.AddEntry("Particle", parent)
        
        child1 := &model.Particle{
                Pdg:    11,
                Vertex: &model.XYZTF{X: 0.5},
                Mass:   0.000511,
                Charge: -3,
        }
        child2 := &model.Particle{
                Pdg:    -11,
                Vertex: &model.XYZTF{X: 0.5},
                Mass:   0.000511,
                Charge: 3,
        }

        kids := event.AddEntries("Particle", child1, child2)
        // tag the stable particles with a special tag
        for _, id := range kids {
                event.TagEntry(id, "GenStable")
        }

        // use assigned identifiers to reference particle lineage
        parent.Child = append(parent.Child, kids...)
        child1.Parent = append(child1.Parent, pid)
        child2.Parent = append(child2.Parent, pid)

        fmt.Println(event)
        // push the event into the stream (save to file)
        writer.Push(event)
}
\end{minted}

\paragraph{Read and Iterate Particles}
\begin{minted}[linenos]{go}
package main

import (
        "fmt"
        "log"

        "github.com/proio-org/go-proio"
        // import data model protobufs anticipated to be read (no dynamic message
        // deserialization available in go *yet*)
        model "github.com/proio-org/go-proio-pb/model/example"
)

func main() {
        // open a stream reader that inputs from "file.proio"
        reader, err := proio.Open("file.proio")
        if err != nil {
                log.Fatal(err)
        }
        defer reader.Close()

        evnum := 0
        // loop over events in stream (in this case a file), specifying a channel
        // buffer of size 10
        for event := range reader.ScanEvents(10) {
                fmt.Printf("EVENT: %v, RUN: %v\n", evnum, string(event.Metadata["run"]))
                // get entry identifiers tagged as 'Particle'
                parts := event.TaggedEntries("Particle")
                fmt.Printf("%v particle(s)...\n", len(parts))
                // loop over particle identifiers
                for i, id := range parts {
                        // deserialize particle entry and cast to known type (switch can be
                        // used here to dynamically determine type from a set of known
                        // possibilities)
                        part := event.GetEntry(id).(*model.Particle)
                        fmt.Printf("%v. PDG Code: %v\n", i, part.Pdg)
                        fmt.Printf("  %v children...\n", len(part.Child))
                        for j, id := range part.Child {
                                child := event.GetEntry(id).(*model.Particle)
                                fmt.Printf("  %v. PDG Code: %v\n", j, child.Pdg)
                        }
                }

                evnum++
        }
}
\end{minted}

\section{C++ Example}
\label{appCPPExample}
The following example uses cpp-proio \cite{cpp-proioRepo} to write and read in the ProIO format.
\paragraph{Write Particles to File or Stream}
\begin{minted}[linenos]{cpp}
#include "proio/writer.h"

// include data model protobuf headers (this example model happens to be
// generated at build time with cpp-proio, as are a few others)
#include "proio/model/example/example.pb.h"

namespace model = proio::model::example;

void write() {
    // create a writer for outputting to "file.proio" - this will close upon
    // destruction
    proio::Writer writer("file.proio");

    // label all future events as "run" = "jpsi0", until this metadata key is
    // overwritten
    writer.PushMetadata("run", "jpsi0");

    proio::Event event;

    // add particles to the empty event and hold onto their assigned
    // identifiers (entries that are added to an event are then owned by the
    // event and will be deleted along with the event)
    auto parent = new model::Particle;
    parent->set_pdg(443);
    auto vertex = parent->mutable_vertex();
    vertex->set_x(1);
    parent->set_mass(3.097);
    uint64_t parent_id = event.AddEntry(parent, "Particle");

    auto child1 = new model::Particle;
    child1->set_pdg(11);
    vertex = child1->mutable_vertex();
    vertex->set_x(0.5);
    child1->set_mass(0.000511);
    child1->set_charge(-3);
    auto child2 = new model::Particle;
    child2->set_pdg(-11);
    vertex = child2->mutable_vertex();
    vertex->set_x(0.5);
    child2->set_mass(0.000511);
    child2->set_charge(3);

    std::vector<uint64_t> child_ids = {event.AddEntry(child1, "Particle"),
                                       event.AddEntry(child2, "Particle")};
    // tag the stable particles with a special tag
    for (auto id : child_ids) event.TagEntry(id, "GenStable");

    // use assigned identifiers to reference particle lineage
    for (auto id : child_ids) parent->add_child(id);
    child1->add_parent(parent_id);
    child2->add_parent(parent_id);

    std::cout << event.String() << std::endl;
    // push the event into the stream (save to file)
    writer.Push(&event);
}
\end{minted}

\paragraph{Read and Iterate Particles}
\begin{minted}[linenos]{cpp}
#include "proio/reader.h"

// include data model protobuf headers anticipated to be used (can omit this if
// performing dynamic reflection)
#include "proio/model/example/example.pb.h"

namespace model = proio::model::example;

void read() {
    // open a stream reader that inputs from "file.proio" - this will close
    // upon destruction
    proio::Reader reader("file.proio");

    int evnum = 0;
    proio::Event event;
    // loop over events in stream (in this case a file), reusing the same Event
    // object each time
    while (reader.Next(&event)) {
        std::cout << "EVENT: " << evnum << ", RUN: " << *event.Metadata()["run"] << std::endl;
        // get entry identifiers tagged as 'Particle'
        auto parts = event.TaggedEntries("Particle");
        std::cout << parts.size() << " particle(s)..." << std::endl;

        // loop over particle identifiers
        for (int i = 0; i < parts.size(); i++) {
            // deserialize particle entry and cast to known type (dynamic_cast
            // can be used to select from a set of known types, or reflection
            // of the generic Message type can be used)
            auto part = static_cast<model::Particle *>(event.GetEntry(parts[i]));
            std::cout << i << ". PDG Code: " << part->pdg() << std::endl;
            std::cout << "  " << part->child_size() << " children..." << std::endl;
            for (int j = 0; j < part->child_size(); j++) {
                auto child = static_cast<model::Particle *>(event.GetEntry(part->child(j)));
                std::cout << "  " << j << ". PDG Code: " << child->pdg() << std::endl;
            }
        }

        evnum++;
    }
}
\end{minted}

\paragraph{Read and Iterate Particles Dynamically}
\begin{minted}[linenos]{cpp}
#include "proio/reader.h"

// data model protobuf headers not included since dynamic reflection will be used

void read_dyn() {
    // open a stream reader that inputs from "file.proio" - this will close
    // upon destruction
    proio::Reader reader("file.proio");

    int evnum = 0;
    proio::Event event;
    // instruct the Event to ignore entry message types known at compile time
    event.UseGeneratedPool(false);
    // loop over events in stream (in this case a file), reusing the same Event
    // object each time
    while (reader.Next(&event)) {
        std::cout << "EVENT: " << evnum << ", RUN: " << *event.Metadata()["run"] << std::endl;
        // get entry identifiers tagged as 'Particle'
        auto parts = event.TaggedEntries("Particle");
        std::cout << parts.size() << " particle(s)..." << std::endl;

        // loop over particle identifiers
        for (int i = 0; i < parts.size(); i++) {
            // deserialize particle entry
            auto part = event.GetEntry(parts[i]);

            // get protobuf descriptor to begin reflection (this is now pure
            // protobuf and not specific to cpp-proio)
            auto part_desc = part->GetDescriptor();
            // get "pdg" field descriptor
            auto pdg_desc = part_desc->FindFieldByName("pdg");
            // get "child" field descriptor
            auto child_desc = part_desc->FindFieldByName("child");
            // get protobuf reflection interface (again, this is pure protobuf)
            auto part_refl = part->GetReflection();
            
            // now use reflection pieces to examine particle entries
            std::cout << i << ". PDG Code: " << part_refl->GetInt32(*part, pdg_desc) << std::endl;
            int n_children = part_refl->FieldSize(*part, child_desc);
            std::cout << "  " << n_children << " children..." << std::endl;
            for (int j = 0; j < n_children; j++) {
                auto child = event.GetEntry(part_refl->GetRepeatedUInt64(*part, child_desc, j));
                std::cout << "  " << j << ". PDG Code: " << part_refl->GetInt32(*child, pdg_desc)
                          << std::endl;
            }
        }

        evnum++;
    }
}
\end{minted}

\section{Java Example}
\label{appJavaExample}
The following example uses java-proio \cite{java-proioRepo} to read the ProIO format.
\paragraph{Read and Iterate Particles}
\begin{minted}[linenos]{java}
import java.util.Collection;
import proio.Event;
import proio.Reader;
import proio.model.Example;

public class Read {
  public static void main(String args[]) {
    try {
      // open a stream reader that inputs from "file.proio"
      Reader reader = new Reader("file.proio");

      int evnum = 0;
      // loop over events in stream
      for (Event event : reader) {
        System.out.println(
            "EVENT: " + evnum + ", RUN: " + event.getMetadata().get("run").toString("UTF8"));
        // get entry identifiers tagged as "Particle"
        Iterable<Long> parts = event.getTaggedEntries("Particle");
        int n_parts = ((Collection<?>) parts).size();
        System.out.println(n_parts + " particle(s)...");

        int i = 0;
        // loop over particle identifiers
        for (Long part_id : parts) {
          // deserialize particle entry and cast to known type
          Example.Particle part = (Example.Particle) event.getEntry(part_id);
          System.out.println(i + ". PDG Code: " + part.getPdg());
          System.out.println("  " + part.getChildCount() + " children...");
          int j = 0;
          for (Long child_id : part.getChildList()) {
            Example.Particle child = (Example.Particle) event.getEntry(child_id);
            System.out.println("  " + j + ". PDG Code: " + child.getPdg());
            j++;
          }
          i++;
        }

        evnum++;
      }

      reader.close();
    } catch (Throwable e) {
      e.printStackTrace();
    }
  }
}
\end{minted}

\end{document}